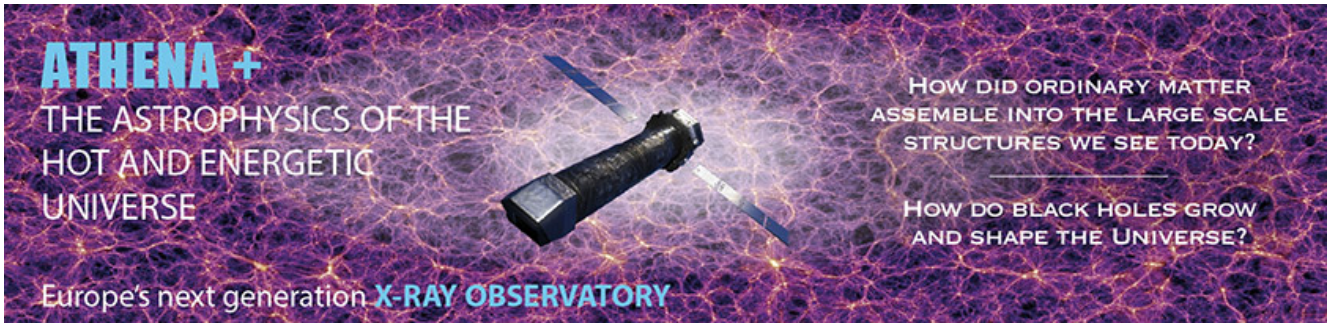

# The Hot and Energetic Universe

An *Athena+* supporting paper

## Understanding the build-up of supermassive black holes and galaxies at the heyday of the Universe

Authors and contributors

**A. Georgakakis, F. Carrera,** G. Lanzuisi, M. Brightman, J. Buchner, J. Aird, M. Page, M. Cappi, J. Afonso, A. Alonso-Herrero, L. Ballo, X. Barcons, M. T. Ceballos, A. Comastri, I. Georgantopoulos, S. Mateos, K. Nandra, D. Rosario, M. Salvato, K. Schawinski, P. Severgnini, C. Vignali

# 1. EXECUTIVE SUMMARY

An outstanding question of current astrophysical research is how galaxies evolve from the peak of the star-formation density of the Universe at redshifts $z$~1-4 to the present time. Observations in the last decade have provided strong evidence that the growth of supermassive black holes at the centres of galaxies is among the most influential processes in galaxy evolution. The generic scenario proposed involves an early phase of intense black hole growth that takes place behind large obscuring columns of inflowing dust and gas clouds. It is postulated that this is followed by a blow-out stage during which some form of AGN feedback controls the fate of the interstellar medium and hence, the evolution of the galaxy.

Open questions that relate to our current understanding of black hole growth and its relation to the build-up of galaxies include: what are the physical conditions (e.g. fuelling mode, triggering mechanism) that initiate major black hole accretion events; what is the nature of AGN feedback and whether it plays a significant role in the evolution of galaxies. X-rays are essential for addressing these points as they uniquely probe AGN at both the early heavily obscured stage and the later blow-out phase.

X-ray spectral analysis can identify the smoking gun evidence of heavily obscured black hole growth (e.g. intense iron K$\alpha$ line). It therefore provides the most robust method for compiling clean samples of deeply shrouded AGN with well-defined selection functions and unbiased determinations of their intrinsic properties (accretion luminosity, obscuring column). X-rays are also the best window for studying in detail AGN feedback. This process ultimately originates in the innermost regions close to the supermassive black hole and is dominated, in terms of energy and mass flux, by highly ionised material that remains invisible at other wavelengths.

The most important epoch for investigating the relation between AGN and galaxies is the redshift range $z$~1-4, when most black holes and stars we see in the present-day Universe were put in place. Unfortunately, exhaustive efforts with current high-energy telescopes only scrape the tip of the iceberg of the most obscured AGN population. Moreover, X-ray studies of the incidence, nature and energetics of AGN feedback are limited to the local Universe.

The *Athena+* mission concept will provide the technological leap required for a breakthrough in our understanding of AGN and galaxy evolution at the heyday of the Universe. Its high throughput will allow the systematic study of AGN feedback to $z$~4 via the identification and measurement of blue-shifted X-ray absorption lines. The excellent survey and spectral capabilities of *Athena+*/WFI (effective area, angular resolution, field of view) will complete the census of black hole growth by yielding samples of up to 100 times larger than is currently possible of the most heavily obscured, including Compton thick, AGN to redshifts $z$~4. The demographics of this population relative to their hosts is fundamental for understanding how major black hole growth events relate to the build-up of galaxies.

## 2. INTRODUCTION

Observational studies in the last 15 years suggest a paradigm shift in our understanding of galaxy formation. The ubiquity of supermassive black holes in local galaxy spheroids has been established (Magorrian et al. 1998). Scaling relations between proxies of the stellar mass of galaxy bulges and the mass of the black holes at their centres have been unexpectedly discovered (e.g. Gebhardt et al. 2000). The cosmological evolution of Active Galactic Nuclei (AGN) and galaxies were shown to follow very similar patterns (e.g. Hopkins & Beacom 2006; Aird et al. 2010). It is therefore argued that the formation of stars in galaxies is related to the growth of supermassive black holes at their centres. In this picture, our knowledge of galaxy evolution will remain incomplete unless we understand this interplay.

The large energy output of AGN relative to the binding energy of their hosts motivated analytical calculations (e.g. Fabian 1999; King 2005) and numerical simulations (e.g. Di Matteo et al. 2005) that proposed AGN feedback as the process that regulates star-formation. The generic picture emerging is that major black hole accretion events and intense star-formation episodes occur nearly simultaneously as the result of gas inflows. The early stages of black hole growth take place behind dense, possibly Compton thick, dust and gas cocoons, which are eventually blown away by some form of AGN outflow. The result is the depletion of the gas reservoirs and the regulation of both star-formation and black hole growth in the galaxy.

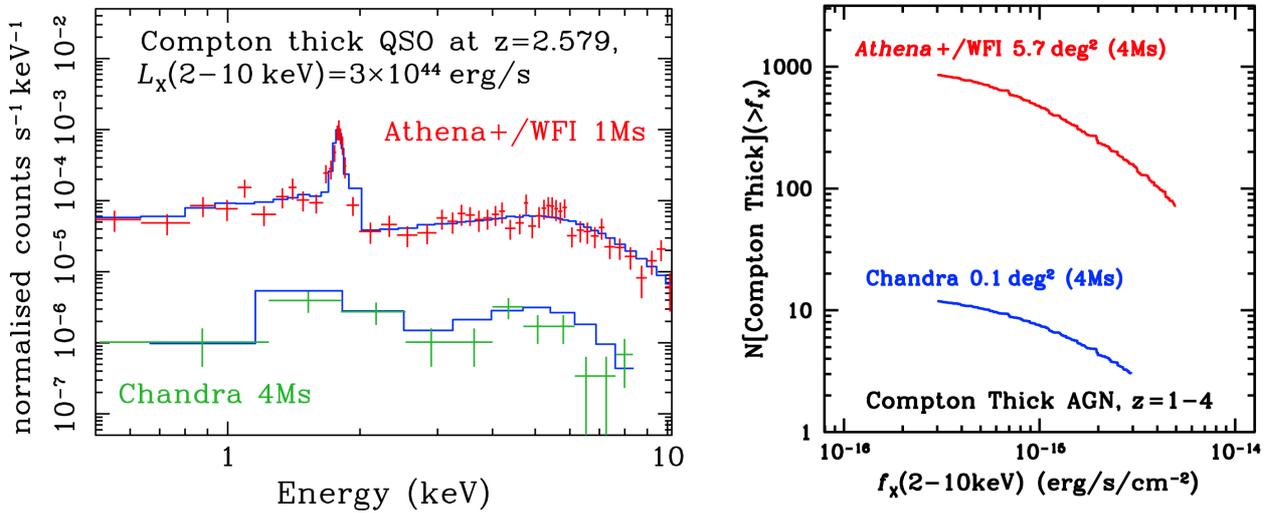

**Fig. 1: Effectiveness of Athena+/WFI surveys for Compton thick AGN population studies:** the plot on the **LEFT** demonstrates the spectral quality that Athena+/WFI surveys will deliver for heavily obscured AGN. The red data-points are the simulated Athena+/WFI 1Ms spectrum of a Compton thick AGN at z=2.579. The most prominent feature in that spectrum is the Fe Ka emission line. The simulated source has been previously identified in the 4Ms Chandra Deep Field South by Brightman & Ueda (2012). The Chandra spectrum is also shown for comparison (green points). The blue lines show the Compton thick spectral model convolved with the Athena+/WFI and Chandra responses. **RIGHT:** Predicted cumulative counts of Compton thick AGN at redshifts z=1–4 as a function of hard (2-10keV) flux. The blue curve corresponds to the 4Ms Chandra Deep Field South, the deepest X-ray image currently available. The red curve is the predicted number counts for an Athena+/WFI survey that consists of 13 pointings of 300ks each, i.e. the same total exposure time as the 4Ms Chandra Deep Field South. Athena+/WFI has the throughput and field-of-view to cover large areas of the sky fast and hence yield, for the same time investment, up to a factor of ~100 more Compton thick AGN compared to the current Chandra X-ray surveys. The calculations assume the X-ray background synthesis model of Akylas et al. (2012; similar results are obtained for the Gilli et al. 2007 model) and the WFI sensitivity for the baseline specifications of 5′′ PSF, 2m$^2$ collecting area at 1keV and 40acrmin$^2$ field-of-view. Confusion is not a problem for 2-10keV sensitivity calculations even for 1Ms WFI exposures. Improvements in the Athena+/WFI parameters toward the goal specifications translate to an increase in the survey speed. For a 3′′ PSF size for example, the WFI will be a factor ~1.5 faster compared to the 5′′ PSF.

In the current paradigm there are two stages that hold important clues on the relation between AGN and galaxies. The early heavily obscured period and the blow-out phase. X-ray observations are a powerful tool for studying both these evolutionary stages, to explore the initial conditions of major black hole growth events and investigate the relevance of AGN feedback in regulating the build-up of galaxies. X-ray imaging/spectroscopy is arguably the cleanest method for finding obscured AGN, even extreme Compton thick ones (Fig. 1), with minimal contamination and a well-defined selection function. Moreover, X-ray observations of nearby Seyferts have revealed outflows that are usually dominated by material that is so highly ionised as to be invisible at other wavelengths and therefore X-rays are the only means by which to detect their imprint and determine their energetics.

The most important period of the Universe to explore AGN/galaxy co-evolution scenarios is the redshift range $z\sim1$-$4$ when the bulk of the stars and supermassive black holes formed. Unfortunately, besides the challenges in characterizing star formation at those redshifts with optical/IR observations, the current deepest X-ray campaigns offer only a glimpse of the obscured active supermassive black hole population at this important epoch. For example, only up to few tens of candidates of the most deeply shrouded Compton thick AGN (e.g. Comastri et al. 2011; Brightman & Ueda 2012; Georgantopoulos et al. 2013) have been unveiled to date in *XMM-Newton* and *Chandra* surveys, while hard X-ray mission like *NuSTAR* will be dominated by Compton thick AGN at $z<1$. Moreover, current or upcoming (e.g. *ASTRO-H*) X-ray telescopes are limited to nearby systems for the study of AGN outflows. A mission like *Athena+* is needed for a systematic study of AGN feedback outside the local Universe and the compilation of large secure samples of the most heavily obscured AGN across redshift.

## 3. OBSCURED ACCRETION AND GALAXY FORMATION

**Multi-tiered surveys with *Athena+*/WFI will identify thousands of the most obscured, including Compton thick, AGN to redshifts $z\sim4$, with well understood selection, minimal contamination and an unbiased determination of their intrinsic properties. The demographics of that population in relation to their hosts is fundamental for understanding the physical conditions under which major black hole growth events at the heyday of the Universe are initiated.**

One of the legacies of *XMM-Newton* and *Chandra* is their survey programs, which allowed population studies of the galaxies experiencing black hole growth. These surveys demonstrated the complexity of the physics that govern the accretion history of the Universe (e.g. Alexander & Hickox 2012). There is evidence for multiple AGN fuelling modes (e.g. cold gas accretion, Di Matteo et al. 2005, vs. hot gas accretion, Croton et al. 2006), diverse triggering mechanisms (e.g. mergers vs. secular processes, Georgakakis et al. 2009, Kocevski et al. 2012), a complex and still not well-understood relation to star-formation in the host galaxy (e.g. Rosario et al. 2012). Moreover, it is fundamental to understand whether the conditions under which black holes grow are a function of cosmic time, large-scale environment, gas availability etc. Demographics are therefore a powerful tool for revealing complex underlying relations between AGN properties (e.g. Eddington rate, accretion luminosity) and diagnostics of physical conditions on larger scales, e.g. star-formation, gas content, morphology, stellar mass, environment of host galaxy (see for example Schawinski et al. 2010, Treister et al. 2012). This type of investigation, however, requires large samples to account for the potentially large intrinsic scatter of relations and covariances between parameters of interest.

Despite the breakthroughs of *XMM-Newton* and *Chandra*, surveys by these missions are limited to moderate numbers, mild obscurations and redshift distributions that peak at $z\sim1$. *Athena+*/WFI has the throughput, field of view and collecting area to complete the census of black hole growth by compiling large AGN samples well into the Compton thick regime and to redshifts $z\sim4$, when the bulk of the black holes we see in present-day galaxies were built.

The most heavily obscured and Compton thick AGN represent a major, although poorly constrained, component of black hole growth (Gilli et al. 2007; Akylas et al. 2012). They are also thought to probe the early phases of major accretion events in galaxies and hence, provide important tests to AGN/galaxy co-evolution models and simulations. Unfortunately, Compton thick AGN are scarce in current surveys (e.g. total of ~40, Brightman & Ueda 2012).

Fig. 1 demonstrates the improvements that *Athena+*/WFI surveys can deliver in detecting and characterising Compton thick AGN. X-ray spectroscopy with the WFI will identify the smoking gun evidence of deeply shrouded black hole growth, e.g. the iron K$\alpha$ line and/or the shape of the X-ray continuum, to yield robust Compton thick identifications for samples up to almost hundred times larger than currently possible for the same time investment. No other high energy mission, currently in operation or planned in the future, can deliver such large numbers of Compton thick AGN at redshifts $z>1$. Moreover, these samples will feature a well-understood selection function, minimal contamination from less obscured sources and an unbiased estimate of the intrinsic AGN power (see Fig. 2).

The *Athena+*/WFI survey strategy (4x1Ms+20x300ks+75x100ks+250x30ks; Aird, Comastri et al., 2013, *Athena+* Support Paper) will yield more than 15,000 Compton Thick AGN with a balanced sampling of both the faint-end (1Ms WFI exposures) and the bright-end of the X-ray luminosity function. Such a large sample over a wide luminosity baseline will allow for the first time a systematic study of the initial conditions of major black hole growth events and an accurate census of the accretion density of the Universe, independent of obscuration biases, to $z\sim4$.

In addition to Compton thick AGN, multi-tiered surveys with *Athena+*/WFI (see survey strategy above) will yield about 180,000 (2-10keV band) mildly obscured AGN in the redshift interval $z=1-4$. This is huge improvement in size compared to current samples (few hundreds) will open a new discovery space in black hole growth investigations, similar to the progress achieved in galaxy evolution studies, when huge datasets (e.g. SDSS) became available (e.g. discovery of star-formation main sequence, galaxy colour bimodality).

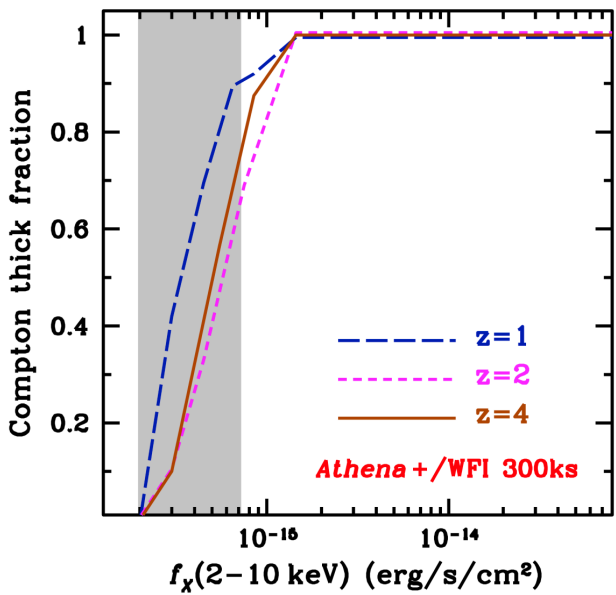

Fig. 2: Athena+/WFI will provide clean Compton thick AGN samples with a well understood selection function: simulated identification rate of Compton thick AGN in Athena+/WFI 300ks surveys as a function of 2-10keV flux. The identification rate is defined as the fraction of simulated Compton thick sources that are recovered as Compton thick by spectral analysis. Different curves correspond to different simulated redshifts. The shaded region shows the range of fluxes for 5-sigma detection. The faint/bright limits of that region are defined as the minimum flux a source should have to be detected (5-sigma) within the area that corresponds to central 10 and 90% respectively of the Athena+/WFI field of view. The contamination rate (i.e. the fraction of mildly obscured AGN erroneously classified as Compton thick) is virtually zero and is not plotted. The intrinsic (i.e. corrected for obscuration) accretion luminosity of Compton thick sources, at all simulated fluxes, is recovered in an unbiased way by Athena+/WFI spectral analysis. For the calculations we use the spectral templates of Brightman & Nandra (2011) to simulate the Athena+/WFI spectra of AGN with a range of 2-10keV fluxes, absorbing column densities ($logN_H$=22.5-25.5 in units of $cm^{-2}$) and redshifts. These are then fitted in XSPEC using standard procedures to infer the column density and intrinsic (i.e corrected for absorption) accretion luminosity.

## 4. COSMIC FEEDBACK

*Athena+* has the capacity to study in a systematic way the incidence and energetics of AGN winds to redshifts $z\sim4$, at the epoch when black hole and galaxy growth peaked and AGN feedback is predicted to be widespread. Observations by *Athena+* will test one of the key assumptions of AGN/galaxy co-evolution models, the significance of winds launched as part of the accretion process in regulating the evolution of galaxies.

Spectroscopy by *XMM-Newton* and *Chandra* revealed outflows in a large fraction of local Seyferts in the form of blue-shifted absorption lines from ionised species (e.g. Crenshaw et al. 2003, Blustin et al. 2005, Cappi 2006, Tombesi et al. 2010, 2013). These observations demonstrated that the bulk of the energy and mass flux in AGN winds is in the form of ionised material that remains invisible at other wavelengths. X-ray observations are therefore the only way to both detect and determine the energetics of those outflows (Cappi, Done et al., 2013, *Athena+* Support Paper).

If AGN winds are responsible for the observed correlations between black hole and galaxy properties, then they should be widespread and energetically significant in the redshift range $z\sim1-4$, when the bulk of black hole growth and star-formation occurred. Unfortunately, outside the local Universe X-ray constraints on AGN outflows are limited to a handful of systems (e.g. Page et al. 2011; Saez & Chartas 2011; Lanzuisi et al. 2012). Upcoming X-ray missions with good spectroscopic capabilities, like *ASTRO-H*, will allow more detailed studies of local sources but will be unable to probe AGN feedback at higher redshifts. It is *Athena+* with its large collecting area and good energy resolution that will open up the opportunity to study in a systematic way AGN winds and outflows out to redshifts $z\sim 4$.

Fig. 3 shows *that Athena+/WFI* has the capacity to identify evidence for outflows in the spectra of AGN and provide rough estimates of their properties (e.g. column density, ionisation state and even velocities for winds at the extremes of the parameter space, Fig. 3-right). Follow-up spectroscopy with *Athena+/X-IFU* of targets selected in *Athena+/WFI* surveys will set tight constraints on the outflow velocities of material over a range of ionisation states and column densities. A dedicated follow-up program of 5Ms with X-IFU will estimate the wind parameters of hundreds of AGN close to the knee ($L^*$) of the X-ray luminosity function, which dominate the growth of black holes at $z\sim 1-4$. These observations, combined with knowledge accumulated from nearby systems (Cappi, Done et al., 2013, *Athena+* Support Paper), will determine the incidence, nature and energetics of AGN outflows for large AGN samples out to $z\sim 4$. This is key to test AGN/galaxy co-evolution scenarios that postulate AGN winds for regulating star-formation.

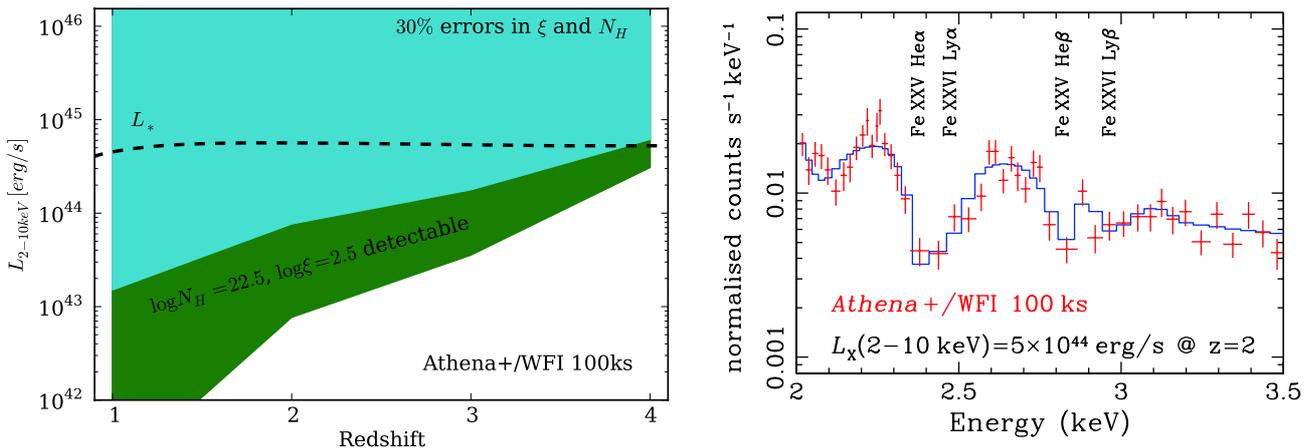

Fig. 3: Power of Athena+/WFI for the identification of AGN outflows: LEFT: AGN $L_X$-z plane over which Athena+/WFI spectroscopy can identify and measure mildly ionized outflows (warm absorbers). A warm absorber model with ionisation log$\xi$=2.5, hydrogen column density $N_H$=5×10$^{22}$cm$^{-2}$ and AGN power-law index $\Gamma$=1.9 is simulated at different redshifts and 2-10keV X-ray luminosities assuming a 100ks Athena+/WFI exposure. For simulated AGN spectra in the green-shaded region the ionised (warm) absorption is preferred (30 times more likely) over cold absorption as the model that describes the observed spectrum. The cyan shaded region shows the $L_X$-z plane over which Athena+/WFI 100ks spectra will in addition measure the parameters of the mildly ionised material ($\xi$, $N_H$) with accuracy better than 30%. These objects can then be targeted by Athena+/X-IFU to measure the velocities and energetics of the outflows. The dashed curve plots the knee (L·) of the X-ray luminosity function (PLE model of Aird et al. 2010). AGN of that luminosity produce most of the accretion density at any given redshift. RIGHT: Simulated Athena+/WFI 100ks spectrum of an AGN at z=2 and $L_X$(2-10keV)=5×10$^{44}$erg/s with a highly ionised (log$\xi$=3.5, $N_H$=10$^{24}$cm$^{-2}$, turbulent velocity 1000km/s) ultra-fast (v=0.2c) outflow. For these extreme systems the Athena+/WFI spectra have sufficient S/N to measure the absorption lines imprinted by ionised species and estimate the outflow velocity, column density and ultimately energy flux. The high throughput of Athena+/WFI, which translates to good S/N in moderate exposures, is key for the identification and study of such ultra-fast high-ionisation winds, which are believed to be highly variable (Lanzuisi et al. 2012). The most efficient way to find and study those winds are surveys with Athena+/WFI.

## 5. SYNERGIES WITH OTHER FACILITIES

*Athena+* will reveal the sites of black hole growth out to high redshift and high obscurations and characterise their properties. A number of large facilities across wavelength planned for the late 2010s and/or 2020s are expected to be an excellent complement to the *Athena+* capabilities.

*JWST* for example will permit detailed near and mid-infrared spectroscopic diagnostics and kinematics of cold gas and dust in star-forming galaxies in the redshift range $z\sim 1-4$ which will have a strong impact on our understanding of galaxy evolution. *SPICA* will extend these spectroscopic measurements into the far infrared. In turn the X-IFU spectrometer on *Athena+* will diagnose the energetic contribution of the black hole both in radiation and in the kinetic power of ionised outflows.

On the other hand *Euclid* is expected to deliver redshifts and spectra of many tens of thousands of galaxies in the range 1<z<3, over the 40deg$^2$ that will constitute the *Euclid* deep surveys. This large statistical sample will have a strong impact on our understanding of galaxy evolution much as SDSS has achieved at low redshifts. *Athena+* surveys carried out within the *Euclid* deep survey areas will provide robust identifications of the galaxies that experience black hole growth to study their environments, stellar masses and host galaxy colours, relative to non-AGN samples.

Similarly, continuum radio surveys by the SKA, which is expected to operate at the same timescale as *Athena+*, will reach flux density depths (nano-Jy) that are currently largely unexplored. The immediate question that those surveys will have to address is what fraction of the nano-Jy radio population is powered by AGN (Norris et al. 2013). *Athena+* is the facility to answer this question. Moreover, HI SKA surveys out to $z\sim4$ will provide information on the cold gas content of the AGN that *Athena+* will detect.

*Athena+* will also identify the most interesting and relevant sources for AGN/galaxy co-evolution studies to be followed by facilities such as E-ELT, *JWST* and ALMA. Observations by those telescopes will e.g. link X-ray detected AGN winds at $z\sim1$-4 to low ionisation (e.g. UV absorption lines) and cold (CO line wings) outflows, or measure the star-formation rates (mid-IR spectroscopy) and host galaxy morphologies (near-IR imaging) of extreme AGN populations (e.g. Compton thick).